\documentclass{PoS}

\title{A Monte Carlo template-based analysis for very high definition imaging atmospheric Cherenkov telescopes as applied to the VERITAS telescope array}

\ShortTitle{A MC template based analysis for the VERITAS IACT array}

\author{\speaker{S.~Vincent} for the VERITAS Collaboration\thanks{veritas.sao.arizona.edu}
	\\
        Deutsches Elektronen-Synchrotron (DESY)\\
        E-mail: \email{stephane.vincent@desy.de}}

\abstract{
We present a sophisticated likelihood reconstruction algorithm for shower-image analysis of imaging Cherenkov telescopes. 
The reconstruction algorithm is based on the comparison of the camera pixel amplitudes with the predictions from a Monte Carlo based model. 
Shower parameters are determined by a maximisation of a likelihood function. 
Maximisation of the likelihood as a function of shower fit parameters is performed using a numerical non-linear optimisation technique. 
A related reconstruction technique has already been developed by the CAT and the H.E.S.S. experiments, and provides a more precise direction 
and energy reconstruction of the photon induced shower compared to the second moment of the camera image analysis.
Examples are shown of the performance of the analysis on simulated gamma-ray data from the VERITAS array.  
}

\FullConference{The 34th International Cosmic Ray Conference,\\
		30 July- 6 August, 2015\\
		The Hague, The Netherlands}

\begin{document}

\section{Introduction}
Earth-bound gamma-ray detectors make use of particle showers generated by gamma-ray interactions at high altitude. 
High-energy gamma rays (as well as hadrons and leptons) enter the Earth's atmosphere and generate a cascade of secondary particles, forming an extended air shower.
For gamma-rays of $0.1-100~\mathrm{TeV}$ few air-shower particles reach ground level. 
However, secondary particles emit Cherenkov radiation resulting in illumination of a $\sim10^{6}~\mathrm{m^2}$ patch of the ground for a few nanoseconds. 
One should be aware that only a fraction of less than $10^{-4}$ of the total shower energy is converted into Cherenkov photons, and quite a few of these photons get lost before hitting the ground.

To reconstruct the direction and energy of the primary gamma ray and to discriminate them from charged cosmic rays most of the current experiments use reconstruction techniques based on the
second moments of the distribution of pixel amplitudes in the camera~\cite{mac83, hil85}. 
These techniques are very robust and efficient, and relatively good angular resolution ($<0.1~\mathrm{deg}$ at 1~TeV with VERITAS)  can be reached with such a reconstruction technique. 
However, significant additional information can be extracted from the recorded images, resulting in improved performance. 

A more sophisticated reconstruction technique has been developed by the CAT experiment~\cite{leb98} and improved for the H.E.S.S. collaboration~\cite{nau09, par14}. 
These methods begin with the creation of a template library that contains the expected shower image for a given set of shower parameters. 
The template library can then be compared to the images recorded in a given event, and, by means of a multi-dimensional fit procedure, the best-fit shower parameters determined.

In this paper we present the ongoing development work on an attempt to improve the performance of VERITAS, by the use of a Monte Carlo simulation based air shower model, combined with 
a ray-tracing telescope simulation. 

\section{Model Generation}

This section describes the generation of the image models, i.e. the prediction of the 
expected Cherenkov light distribution in the camera focal plane for a given set of 
primary particle parameters. 
The production of mean shower images is divided into two steps: the generation of large Monte Carlo datasets 
and the simulation of the detector response. 
 
The electromagnetic air showers in the atmosphere are simulated 
with the program CORSIKA~\cite{hec98}. 
The simulations are performed over a range of energy, zenith angle, impact distance, and 
primary interaction depth. 
The positions of the telescopes on the ground are on a plane taken perpendicular to the shower axis, this 
allows for the projection and the camera planes to be parallel.  
The output of the CORSIKA simulations consists of the prediction of the light distribution 
on the ground plus the arrival direction of each photon. 
For each event, the Cherenkov photons falling onto the mirror elements are followed individually 
according to their arrival times, initial direction, and wavelength with 
additional use of the atmospheric density profile, optical absorption and some of the detector 
characteristics such as its light-collecting area, photo-tube quantum efficiency.  

The images are produced for a VERITAS camera with pixels of $0.15^{\circ}$ diameter, and the telescopes use a Davies-Cotton optical design. 
The shower images are generated for a source at the centre of the camera (i.e. $0^{\circ}$ wobble-offset) and the gamma-ray longitudinal 
development is oriented along the X-axis of the camera frame. 
Comparison with offset relative to the camera centre and/or off-axis events will result in a rotation and a translation 
in the camera frame; these transformations are applied later in the fit procedure. 

\begin{figure}[ht]
  \centering
  \begin{minipage}[c]{0.32\textwidth}
    \centering
    \includegraphics[width=\textwidth]{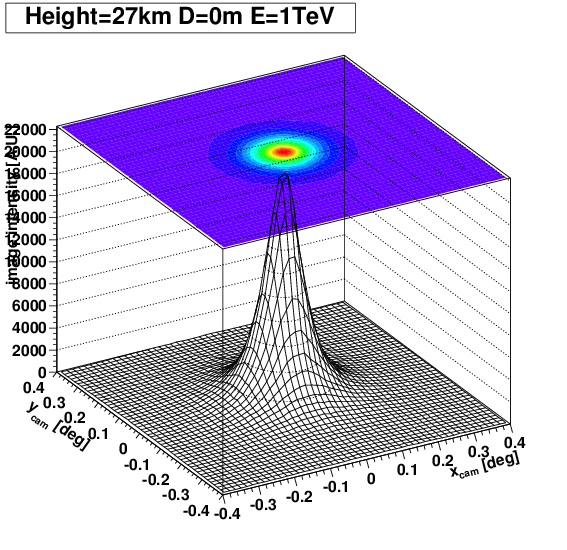}
  \end{minipage}
  \begin{minipage}[c]{0.32\textwidth}
    \centering
    \includegraphics[width=\textwidth]{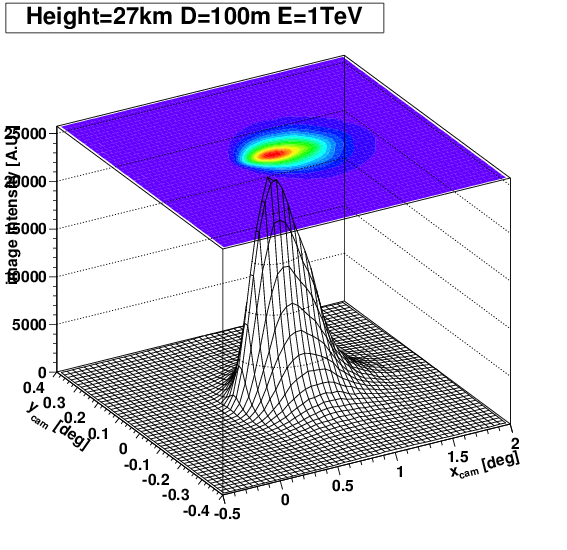}
  \end{minipage}
  \begin{minipage}[c]{0.32\textwidth}
    \centering
    \includegraphics[width=\textwidth]{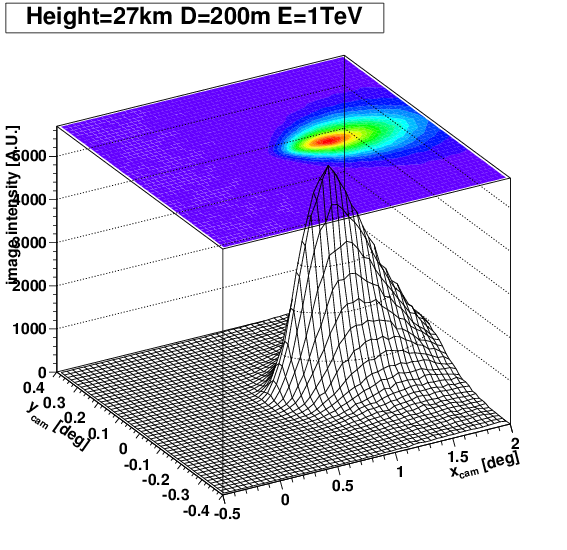}
  \end{minipage}
  \caption{Image template histograms for a 1~TeV primary gamma-ray at a core distance of 0~m (left), 100~m (centre) and 200~m (right) at first interaction depth of 27~km. 
  The x and y axes are in degrees. The z-axis is given in arbitrary units.}
\label{images_D}
\end{figure}

The models are generated for 9 first-interaction depths, from 0$\cdot X_{0}$ and 5$\cdot X_{0}$ (the radiation length in air is $X_{0}=36.7~ \mathrm{g.cm}^{-2}$), 90 energies, from 30 GeV 
to 30 TeV, and 50 impact distances, from 0 to 500 m. 
A multidimensional interpolation algorithm is used to interpolate between the templates, allowing production of an image template 
for any shower parameters within the parameter ranges.
Examples of bi-dimensional profiles of 1 TeV shower images obtained from Monte-Carlo simulations 
are shown in Fig. \ref{images_D} for different values of the impact parameter. 

\begin{figure}[ht]
  \centering
  \begin{minipage}[c]{0.49\textwidth}
    \centering
    \includegraphics[width=\textwidth]{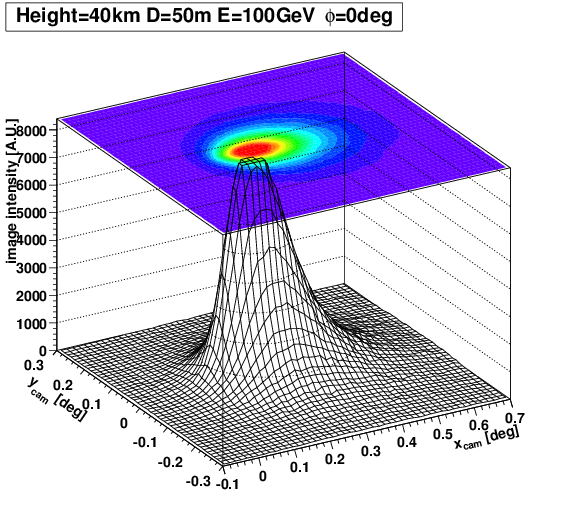}
  \end{minipage}
  \begin{minipage}[c]{0.49\textwidth}
    \centering
    \includegraphics[width=\textwidth]{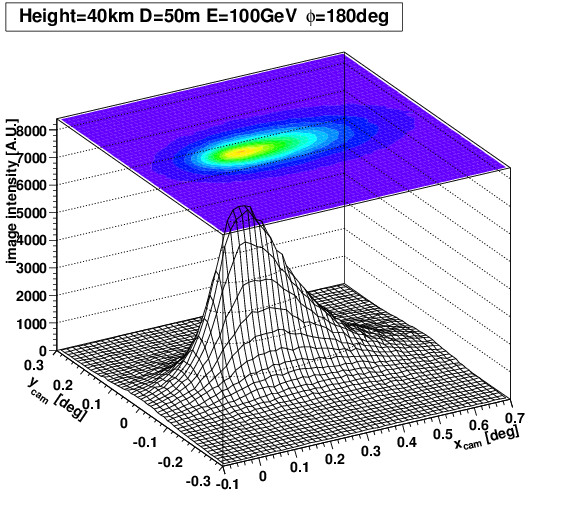}
  \end{minipage}
  \caption{Image template for two values of the azimuthal angle of the shower axis direction: $\phi=0^{\circ}$, $\phi=180^{\circ}$. 
  The templates shown are evaluated for a gamma-ray shower with an energy of 100 GeV, an impact distance of 50 m, and a first interaction depth of 40 km. 
   Note that the vertical scale (image amplitude) is given in arbitrary units but the same number of air showers were simulated. 
    The geomagnetic field affects the development of showers. 
    The images are diffused and distorted when the component of the field is normal to the shower. }
\label{images_Az}
\end{figure}

An additional parameter included here is the effect of the geomagnetic field on the electromagnetic showers. 
We treat the geomagnetic field $\mathbf{B}$ at the VERITAS location,
\begin{equation}
B=47~\mu\mathrm{T} \quad D=10^{\circ}21' \quad I=58^{\circ}13' 
\end{equation}
$D$ and $I$ being the geomagnetic declination and inclination. 
The declination $D$ is the angle between the magnetic field and the true north. 
The inclination $I$ is the angle between the magnetic field vector and the horizontal plane that is tangent to the Earth's surface at the observer's position. 
Fig.~\ref{images_Az} shows simulated camera images for two values of the shower direction. 

For the present, the position of the gamma-ray source within the camera field of view is ignored. 
This could be important due to the broadening of the telescopes optical point spread function with distance from the camera centre. 

\section{Likelihood Reconstruction \& Performance}
Once the full set of templates has been created, they must be compared with the observed images. 
The likelihood reconstruction performs a global fit to the telescope image data using a model for the expected pixel amplitude. 
Shower parameters are determined by a maximisation of an array likelihood function developed in~\cite{nau09}. 
The maximisation of this array likelihood is performed using a numerical non-linear optimisation technique. 

The probability density provides a model of a measured pixel signal $s$ in units of photoelectrons (PE) given an expectation of $\mu$, and consists of a convolution 
of the Poisson distribution of the photoelectron number~$n$, with the resolution of the photosensor. 
The latter is well represented by a Gaussian of width $\sqrt{\sigma^{2}_{\mathrm{p}}+n\sigma^{2}_{\gamma}}$, 
where $\sigma_{\mathrm{p}}$ is the standard deviation of the detector noise and $\sigma_{\gamma}$ is the width of the single PE response function~\cite{han09}:
\begin{equation}
P(s|\mu, \sigma_{\gamma}, \sigma_{\gamma})=\sum_{n}\frac{\mu^n e^{-\mu}}{n!\sqrt{\sigma^{2}_{\mathrm{p}}+n\sigma^{2}_{\gamma}}}
\mathrm{exp}\left(-\frac{(s-n)^2}{2(\sigma^{2}_{\mathrm{p}}+n\sigma^{2}_{\gamma})}\right)
\end{equation}
The pixel log-likelihood $L_{\mathrm{pix}}$ function is 
\begin{equation}
\mathrm{ln}~L_{\mathrm{pix}}=-2\times\mathrm{ln}~P(s|\mu, \sigma_{\mathrm{p}}, \sigma_{\gamma}).
\end{equation}
The array likelihood $L_{\mathrm{array}}$ is calculated from the sum over the pixel likelihoods in all telescopes, 
\begin{equation}
\mathrm{ln}~L_{\mathrm{array}}=\sum_{\mathrm{tel.}}\sum_{\mathrm{pix.}}-2\times\mathrm{ln}~P(s|\mu, \sigma_{\mathrm{p}}, \sigma_{\gamma})
\end{equation}
This value is constructed under the assumption 
that if all pixels behave like independent random variables.
The model photon reconstruction relies on the pixel-per-pixel comparison of the actual shower images with the ones that are predicted for a given set of parameters. 
With such a large parameter space, using an appropriate seed position for the fit is crucial to avoid getting trapped in a local minimum. 

The shower reconstruction is performed in two consecutive stages.
The standard moment-based parameters are calculated and used to define the starting point of the minimisation. 
A differential evolution (DE) algorithm may also be used to derive a handful of possible estimates~\cite{pri96, sto96}. 
DE is a stochastic, population-based optimisation algorithm for solving nonlinear optimisation problems. 
It does not require derivatives of the function and an initial population of values is created randomly in a region. 
Eventually, the members of the population converge to the point of highest likelihood value. 
The estimate that provides the best initial log-likelihood from DE or moment-based algorithms is used as starting point of the minimisation. 

The array likelihood must be minimised in a 6-dimensional fit over direction ($x_{\mathrm{target}}, y_{\mathrm{target}}$), 
impact parameter ($x_{\mathrm{imp}}, y_{\mathrm{imp}}$), primary energy, and first interaction depth $X_{0}$. 
Fitting is performed using a Levenberg-Marquardt (LM) algorithm. 
This is a standard technique used to solve nonlinear least squares problems. 
It is actually a combination of two minimisation methods: the gradient descent method and the Gauss-Newton method. 
The iteration continues until the convergence criteria or the stopping criteria are satisfied. 
Convergence is assumed if either the state vector does not change by more than a set value between iterations
or the cost function does not change by more than a set value between iterations. 
We also provide a stopping criteria for when a maximum number of iterations of
the algorithm is reached. 
The outputs of the minimisation procedure are the 6 shower parameters, the correlation matrix of the
fit parameters, and the uncertainties on the best fit parameters.

A goodness-of-fit test addresses the question as to how well the data are described by the image template and the test only involves one hypothesis.
To quantify the agreement between the measured and the expected pixel signals, we define a $\chi^2$-like parameter 
\begin{equation}
G=\frac{\mathrm{ln}~L_{\mathrm{array}}-\langle \mathrm{ln}~L_{\mathrm{array}} \rangle}{\sqrt{2\times\mathrm{NdF}}},
\end{equation}
where the average array log-likelihood for a given $\mu$, $\sigma_{\mathrm{p}}$ and $\sigma_{\gamma}$ can be calculated 
as below:
\begin{equation}
\langle \mathrm{ln}~L_{\mathrm{array}} \rangle =\sum_{\mathrm{tel.}}\sum_{\mathrm{pix.}}\int ds \mathrm~\mathrm{ln}~L_{\mathrm{pix}} \times P(s|\mu, \sigma_{\mathrm{p}}, \sigma_{\gamma})
\end{equation}

\begin{figure}[ht!]
  \centering
  \begin{minipage}[c]{0.49\textwidth}
    \centering
    \includegraphics[width=\textwidth]{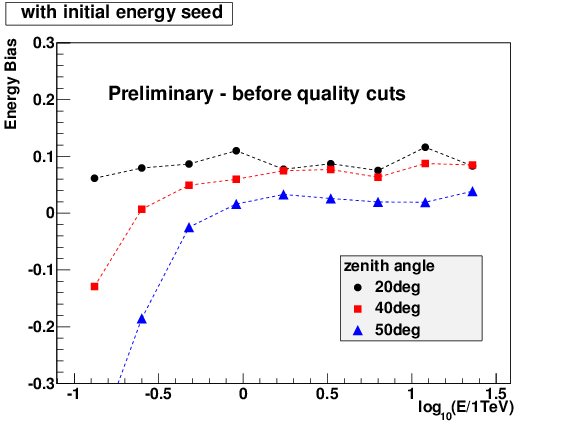}
  \end{minipage}
  \begin{minipage}[c]{0.49\textwidth}
    \centering
    \includegraphics[width=\textwidth]{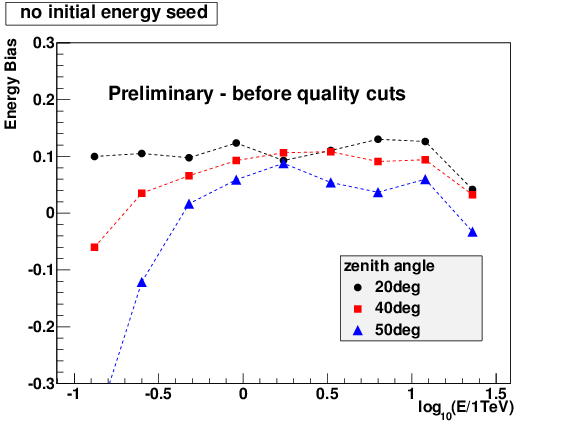}
  \end{minipage}
  \begin{minipage}[c]{0.49\textwidth}
    \centering
    \includegraphics[width=\textwidth]{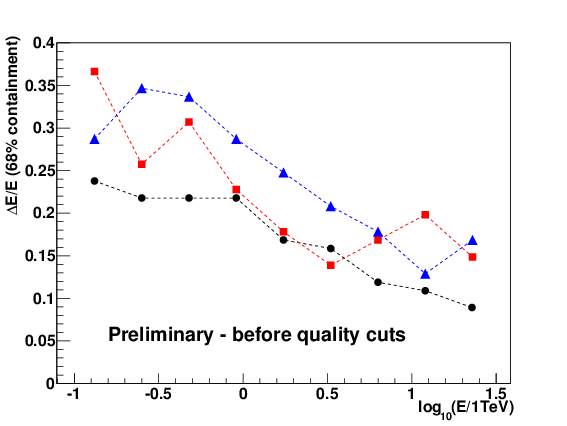}
  \end{minipage}
  \begin{minipage}[c]{0.49\textwidth}
    \centering
    \includegraphics[width=\textwidth]{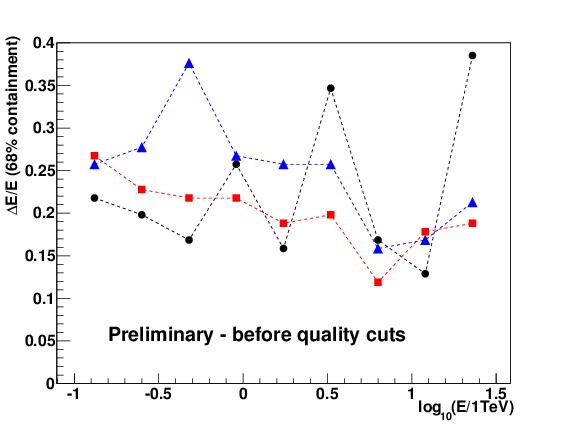}
  \end{minipage}
  \begin{minipage}[c]{0.49\textwidth}
    \centering
    \includegraphics[width=\textwidth]{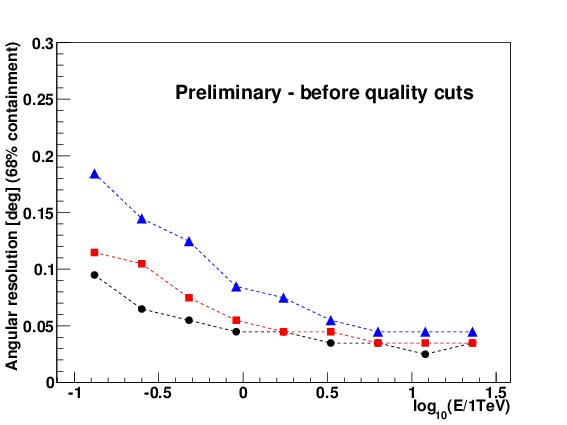}
  \end{minipage}
  \begin{minipage}[c]{0.49\textwidth}
    \centering
    \includegraphics[width=\textwidth]{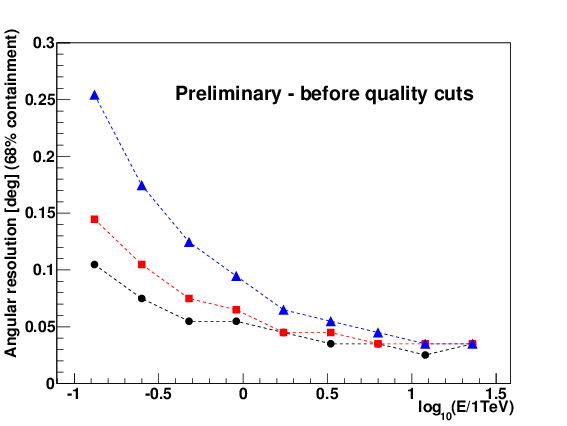}
  \end{minipage}
  \caption{Energy bias $(\mathrm{E}-\mathrm{E}_{\mathrm{rec.}})/\mathrm{E}$ (top), energy resolution (middle), and angular resolution (68\% containment radius) shown as function of the simulated photon energy~$\mathrm{E}$. 
  The energy resolution is defined as the 68\% containment radius around the median value of the energy bias distribution. 
  The left (resp. right) column shows results when the standard Hillas reconstruction technique provides (resp. does not provide) a starting value for the energy of the primary.}
\label{energy_resolution_and_bias}
\end{figure}

The performance is assessed using Monte Carlo simulations. 
We consider events that satisfied the convergence criteria and no quality cuts are applied to the data. 
The energy bias and the energy resolution as function of the simulated energy and zenith angle are shown in Fig.~\ref{energy_resolution_and_bias}. 
The energy bias is defined as the mean of the $\Delta E/E$ distribution. 
The energy resolution is defined as the 68\% containment of the $\Delta E/E$ distribution. 
At low zenith angle, the energy bias changes little in the whole energy range while, at low energies and at large zenith angle, the curves show a large bias. 

The angular resolution is defined as the 68\% containment radius of the reconstructed event positions from a point-like source. 
In contrast to the moment analysis, where the reduction of performance at large zenith angle is due to the larger fraction of nearly parallel images, the template analysis considers 
more information than just the image axis so it should cope better at low elevation.   

\newpage

\section{Summary}
In this contribution, we present the progress of a likelihood-based reconstruction method for the VERITAS telescope array. 
The algorithm is based on an accurate pixel-by-pixel comparison of observed intensities with a Monte Carlo based template. 
This technique has been proven in the past to be extremely successful, reconstructing events with more accuracy than the traditional moment analysis. 
As many fitting algorithms, the LM algorithm has some drawbacks, such as getting stuck in a local minimum or being dependent of the initial parameter values. 
To prevent these problems, we introduce a global minimisation algorithm (differential evolution scheme). 
Its performance is shown in~Fig.\ref{energy_resolution_and_bias}. 

The intrinsic capabilities of the moment analysis and the template analysis (and in particular the hadronic rejection capabilities) can be combined together to improve the sensitivity of the analysis. 
Since these analyses perform differently in different energy and impact parameter domains, more detailed studies should also allow one to select the optimal response on an event-by-event basis and 
therefore improve the quality (angular resolution,...) of the analysis.

Additionally, the template method can also be used to reconstruct any particle type, in particular iron nuclei~\cite{fle15}.

\section{Acknowledgments}
This research is supported by grants from the U.S. Department of Energy Office of Science,
the U.S. National Science Foundation and the Smithsonian Institution, and by NSERC in Canada.
S.V. acknowledges support through the Helmholtz Alliance for Astroparticle Physics.
We acknowledge the excellent work of the technical support staff at the Fred Lawrence
Whipple Observatory and at the collaborating institutions in the construction and operation of the
instrument. The VERITAS Collaboration is grateful to Trevor Weekes for his seminal contributions
and leadership in the field of VHE gamma-ray astrophysics, which made this study possible.

\end{document}